\documentclass[conference, 10pt]{IEEEtran}

\usepackage{cite}
\usepackage{url}
\usepackage{graphicx}
\usepackage{subcaption}
\usepackage{color}
\usepackage{placeins}
\usepackage{float}
\usepackage{tabularx,colortbl}
\usepackage{ifthen}
\usepackage{amsmath}

\makeatletter

\newcounter{author}
\renewcommand{\author}[2][]{
   \stepcounter{author}
   \@namedef{author@\theauthor}{#2}
   \@namedef{authorlabel@\theauthor}{#1}
}

\newcounter{address}
\newcommand{\address}[2][]{
   \stepcounter{address}
   \@namedef{address@\theaddress}{#2}
   \@namedef{addresslabel@\theaddress}{#1}
}

\newcommand{\alsep}{and}

\def\newmaketitle{\par%
  \begingroup%
  \normalfont%
  \def\thefootnote{}%  the \thanks{} mark type is empty
  \def\footnotemark{}% and kill space from \thanks within author
  \let\@makefnmark\relax% V1.7, must *really* kill footnotemark to remove all \textsuperscript spacing as well.
  \footnotesize%       equal spacing between thanks lines
  \footnotesep 0.7\baselineskip%see global setting of \footnotesep for more info
  \normalsize%
  \twocolumn[\thenewmaketitle\@IEEEaftertitletext]%
  \if@IEEEusingpubid
     \enlargethispage{-\@IEEEpubidpullup}%
  \fi
  \endgroup
  \setcounter{footnote}{0}\let\maketitle\relax\let\@maketitle\relax
  \gdef\@thanks{}%
  \let\thanks\relax}

\def\thenewmaketitle{
  \newpage
  \begin{center}%
    \vskip0.2em{\Huge\@IEEEcompsoconly{\sffamily}\@IEEEcompsocconfonly{\normalfont\normalsize\vskip 2\@IEEEnormalsizeunitybaselineskip
   \bfseries\large}\@title\par}\vskip1.0em\par%
    \vspace{1ex}
    \newcounter{c@author}
    \newcounter{c@tmp}
    \ifthenelse{\value{author}=2}{%
      \newcommand{\liand}{ and }}{%
      \newcommand{\liand}{, and }}
    \ifthenelse{\value{address}<2}{%
      \@nameuse{author@1}%
      \stepcounter{c@author}%
      \whiledo{\value{c@author}<\value{author}}{%
        \setcounter{c@tmp}{\value{author}}%
        \addtocounter{c@tmp}{-\value{c@author}}%
        \ifthenelse{\value{c@tmp}=1}{%
          \renewcommand{\alsep}{\liand}}{\renewcommand{\alsep}{, }}%
        \stepcounter{c@author}\alsep \@nameuse{author@\thec@author}}\\%
    }
    {%Add address references after the author's name
      \@nameuse{author@1}${}^{(\ref{\@nameuse{authorlabel@1}})}$%
      \stepcounter{c@author}%
      \whiledo{\value{c@author}<\value{author}}{%
      \setcounter{c@tmp}{\value{author}}%
      \addtocounter{c@tmp}{-\value{c@author}}%
      \ifthenelse{\value{c@tmp}=1}{%
        \renewcommand{\alsep}{\liand}}{\renewcommand{\alsep}{, }}%
      \stepcounter{c@author}\alsep \@nameuse{author@\thec@author}%
        ${}^{(\ref{\@nameuse{authorlabel@\thec@author}})}$%
      }
    }
    \vspace{0.2ex}

    \ifthenelse{\value{address}>0}{%
      \ifthenelse{\value{address}=1}{
        {\@nameuse{address@1}}
      }
      {%Output the addresses as an enumerated list
        \newcounter{c@address}

        \begin{center}
        \whiledo{\value{c@address}<\value{address}}
        {
          \refstepcounter{c@address}
            ${}^{(\thec@address)}$\,%
              \label{\@nameuse{addresslabel@\thec@address}}%
              \@nameuse{address@\thec@address}\\ %
        }
        \end{center}
      } % end of the address creation ifthenelse block
    }
    {
      \relax
    }
  \end{center}
}

\makeatother

\title{Fixed-Frequency Reconfigurable Leaky-Wave Antennas with Simplified Biasing}

\author[org1]{Sambong Jang}
\author[org1]{Minseok Kim$^*$}

\address[org1]{School of Electronic and Electrical Engineering, Hongik University, 94 Wausan-ro, Mapo-gu, Seoul, 121-791, Korea, minseok.kim@hongik.ac.kr}

\begin{document}

\newmaketitle

\begin{abstract}
This work introduces a reconfigurable leaky-waveguide antenna with a simplified biasing scheme for dynamic beam steering at a fixed frequency. Unlike prior metasurface-aided leaky-waveguide antennas that employ tunable metasurfaces as radiative apertures, our approach utilizes them as a waveguide wall solely to control the guided mode. As a result, the proposed structure eliminates the need for local biasing schemes commonly required in earlier designs, significantly simplifying the biasing process. The radiation is achieved by employing passive, angle-independent metasurfaces that allow beam steering across broadside. The feasibility and effectiveness of the proposed design are validated through full-wave simulations.

\end{abstract}

%A few notable examples include those that utilize tunable composite right/left handed meta-atoms, which allow steering of a beam at a fixed operating frequency. By balancing the resonance frequencies of the series and shunt branches of their meta-atoms, these LWAs could overcome the limitations imposed by the open-stopband effect and demonstrated steering of a beam through broadside. Additionally, Kim \textit{et al.} have ....

\section{Introduction}
Since their introduction, leaky-waveguide antennas (LWAs) have been a topic of prominent interest due to their number of inherent advantages, such as a low profile, high directivity, and frequency-dependent beam-scanning capabilities~\cite{Karmokar2020IEEE}. Recently, this interest has been further intensified by advancements in metasurfaces, which have facilitated the development of metasurface-aided LWAs that effectively addressed several longstanding challenges in conventional LWAs. A few notable metasurface-aided LWAs include those utilizing: (i) tunable composite right/left-handed metasurfaces, which overcome the open-stopband effect and enable beam steering through broadside~\cite{Caloz2008IEEE,Damm2010IEEE}; (ii) switchable metasurfaces with `ON'and `OFF' states, allowing dynamic control of the phase constant of the $n^{\text{th}}$ space harmonic coupled to the radiated field~\cite{Karmokar2013IEEE,Li2019IEEE}; and (iii) Huygens' metasurfaces, which offer independent control over both the amplitude and phase of the radiated field, allowing for the generation of complex radiation patterns beyond simple beam steering~\cite{abdo2018IEEE,Kim2021PRApp}.

Regardless of the specific type of metasurfaces employed in metasurface-aided LWAs, their fundamental operating principle remains largely consistent: metasurfaces act as reconfigurable radiative apertures, dynamically converting guided modes into desired radiation. This functionality is made possible by the underlying structure of the metasurfaces, which consist of meta-atoms equipped with active circuit components, such as varactors and PIN diodes. However, this operating principle and design architecture introduce challenges that hinder their practical implementation and scalability. Specifically, steering a beam or shaping the radiation pattern typically requires inhomogeneous surface properties~\cite{Imbert2015IEEE}. As such, the local interaction between a guided mode and meta-atoms has to be precisely controlled, which in turn necessitates individual biasing of each meta-atom (i.e., a \textit{local} biasing). Therefore, complex and cumbersome biasing networks are needed that become increasingly difficult to manage as the systems scale.

%our approach uses them to modulate the guided mode only. In other words, 
%with its scattering properties engineered to be independent of the propagation angle (i.e., the incident angle). These passive surfaces then 

%This streamlined and efficient approach makes the proposed metasurface-aided LWA particularly suitable for applications in reconfigurable wireless communication systems, spaceborne platforms with stringent power and weight constraints, and compact radar systems requiring precise and dynamic beam steering.

To address this limitation, we propose a metasurface-aided LWA that achieves dynamic beam steering at a fixed operating frequency while significantly simplifying the biasing scheme. Unlike previous designs that employ tunable metasurfaces as radiative apertures, our tunable metasurface does not radiate any fields but only modulate the guided mode within the LWA. This key distinction allows the proposed design to operate with \textit{global} biasing (i.e., uniform biasing), entirely eliminating the need for the \textit{local} biasing schemes required in earlier metasurface-aided LWAs. Meanwhile, the radiative aperture in our design remains entirely passive. Specifically, it consists of a pair of passive angle-independent metasurfaces that directly transform the guided mode into a highly directive beam, with its direction controlled by adjusting the propagation angle of the guided mode. The proposed metasurface-aided LWA is well-suited for applications where ease of fabrication and simplified biasing are critical, ranging from wireless communication networks to sensing and imaging systems.

%The main objective of the proposed work is to design an LWA to achieve dynamic beam steering with a simple biasing network. To begin with, we first envision a tunable reflective metasurface to change the incident angle inside the LWA by altering the surface impedance value under the guidance resonance condition in [].

\section{Theory and Concept}
\begin{figure}[t]
    \centerline{\includegraphics[width=\columnwidth]{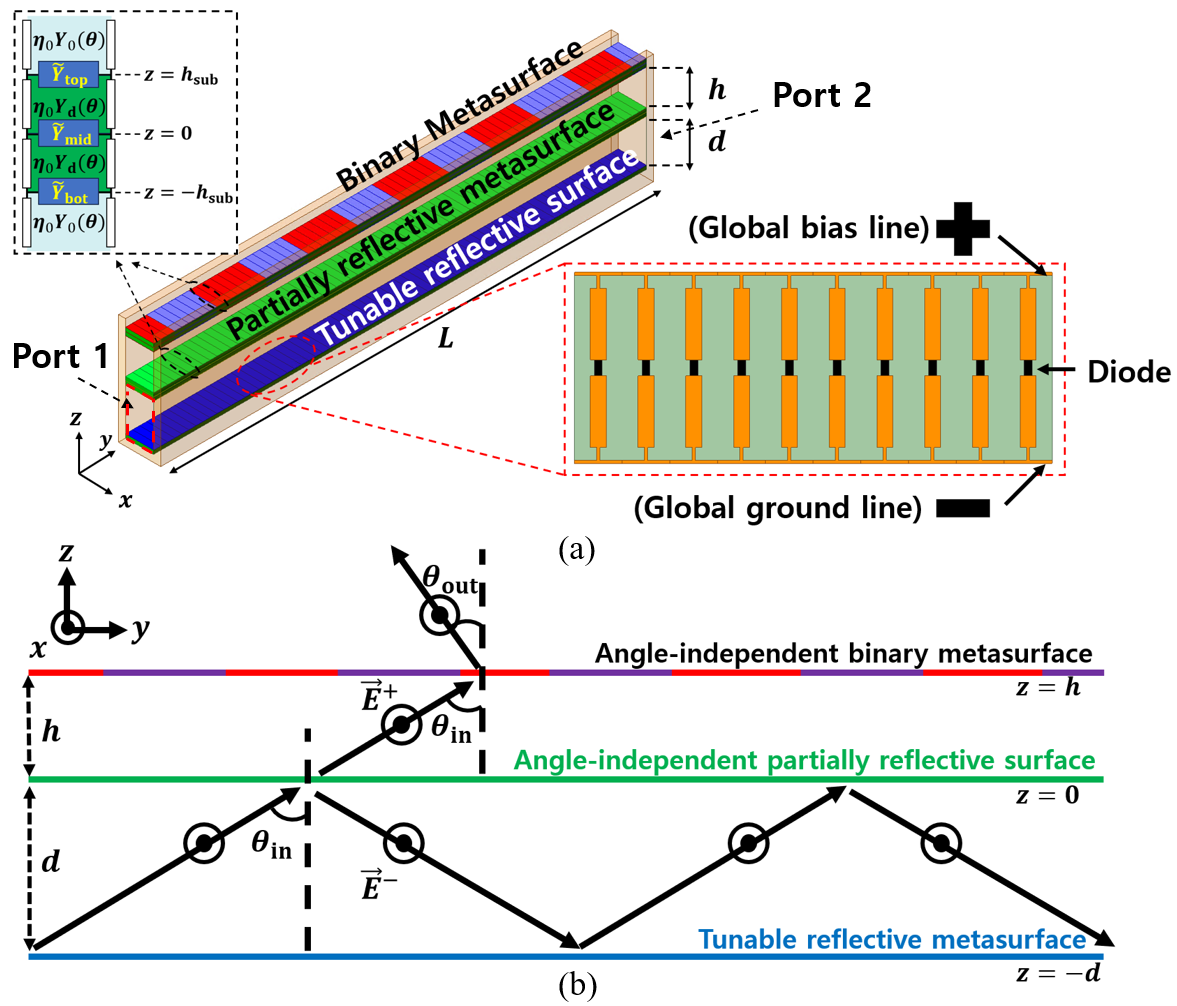}}
    \caption{A schematic of the proposed reconfigurable LWA with simplified biasing. (a) Overview of the proposed reconfigurable LWA design comprising tunable reflective metasurface (TRM), angle-independent partially-reflective surface (PRS), and angle-independent binary metasurface. (b) Cross-section of the proposed LWA.}
    \label{fig:ConceptLWAFig.PNG}
\end{figure}

Fig.\ref{fig:ConceptLWAFig.PNG} illustrates the schematic of the proposed metasurface-aided reconfigurable LWA, which consists of three main components: (i) a tunable reflective metasurface (TRM), (ii) an angle-independent partially reflective surface (PRS), and (iii) an angle-independent binary metasurface (BMS).
Before detailing the underlying principle of the proposed LWA, it is important to note that both the TRM and PRS are homogeneous surfaces, meaning their surface properties remain uniform across their entire area. For the TRM, this homogeneity physically implies that its unit cells are connected in parallel and uniformly biased. This uniform biasing can be easily implemented using global bias lines along the waveguide direction, $y$, as depicted in the inset of Fig. \ref{fig:ConceptLWAFig.PNG}(a). On the other hand, the PRS is modeled as a cascade of three admittance surfaces ($Y_\text{top}$, $Y_\text{mid}$, and $Y_\text{bot}$), separated by a 1.575-mm thick Rogers 5880 substrate ($\epsilon_r=2.2, \delta=0.0009$). Each of these admittance surfaces is assumed to comprise entirely passive and identical unit cells to ensure homogeneity.

Hereafter, we denote the reflection coefficients of TRM and PRS as $\Gamma_{\text{TRM}}$ and $\Gamma_{\text{PRS}}$, respectively. A guided mode can then be formed between the TRM and PRS, provided that the guidance resonance condition is satisfied as,
\begin{equation}
    \angle \Gamma_{\text{TRM}} + \angle \Gamma_{\text{PRS}} - 2k_od\text{cos}(\theta_{\text{in}})=0
    \label{eq:GuidanceCond}
\end{equation}
\noindent where $\theta_{\text{in}}$, $k_o$, and $d$ represents the propagation angle of the guided mode, the wavenumber in free space, and the distance between the TRM and PRS, respectively (see Fig.~\ref{fig:ConceptLWAFig.PNG}(b)). 
\begin{table}[h!]
    \centering
    \begin{tabular}{|c|c|c|c|c|c|c|}
        \hline
        $\theta_\text{in}$ & $20^\circ$ & $30^\circ$ & $40^\circ$ & $50^\circ$ & $60^\circ$ & $70^\circ$  \\ \hline
        $\angle\Gamma_\text{TRM}$   & $-66.4^\circ$  & $-104^\circ$ & $-154^\circ$ & $144^\circ$ & $72^\circ$ & $-7.6^\circ$    \\ \hline
    \end{tabular}
    \caption{Summary of the required reflective phase with respect to the propagation angle of the guided mode.}
    \label{tab:ReflPhaseOfTRM}
\end{table}

%Equivalently, the longitudinal component of the phase constant (i.e. $\beta_x=k_o\text{sin}(\theta_{\text{in}})$) is altered by changes in the bias voltage.

From Eq.~\eqref{eq:GuidanceCond}, it is seen that $\theta_{\text{in}}$ can be dynamically tuned as a function of $\angle \Gamma_{\text{TRM}}$, and Table~\ref{tab:ReflPhaseOfTRM} summarizes the required phase of $\Gamma_{\text{TRM}}$ for synthesizing various $\theta_{\text{in}}$ when $d=0.5\lambda_o$. It is noted that variations in the bias voltage induce adjustments in the reflection phase of TRM, which subsequently modifies the propagation angle in the LWA. Furthermore, if the magnitude of $\Gamma_{\text{PRS}}$ is kept below unity, the PRS would radiate a beam at $\theta_{\text{in}}$ (i.e., the fundamental-mode radiation), which varies as $\angle \Gamma_{\text{TRM}}$ changes according to Eq.~\eqref{eq:GuidanceCond}. Thus, the combination of the TRM and PRS facilitates beam steering with a simple biasing mechanism. However, the reflection properties of the PRS are not guaranteed to remain independent of $\theta_{\text{in}}$. Moreover, this method fails to achieve broadside beam steering as the fundamental mode is radiated.

\begin{figure}
    \centerline{\includegraphics[width=0.8\columnwidth]{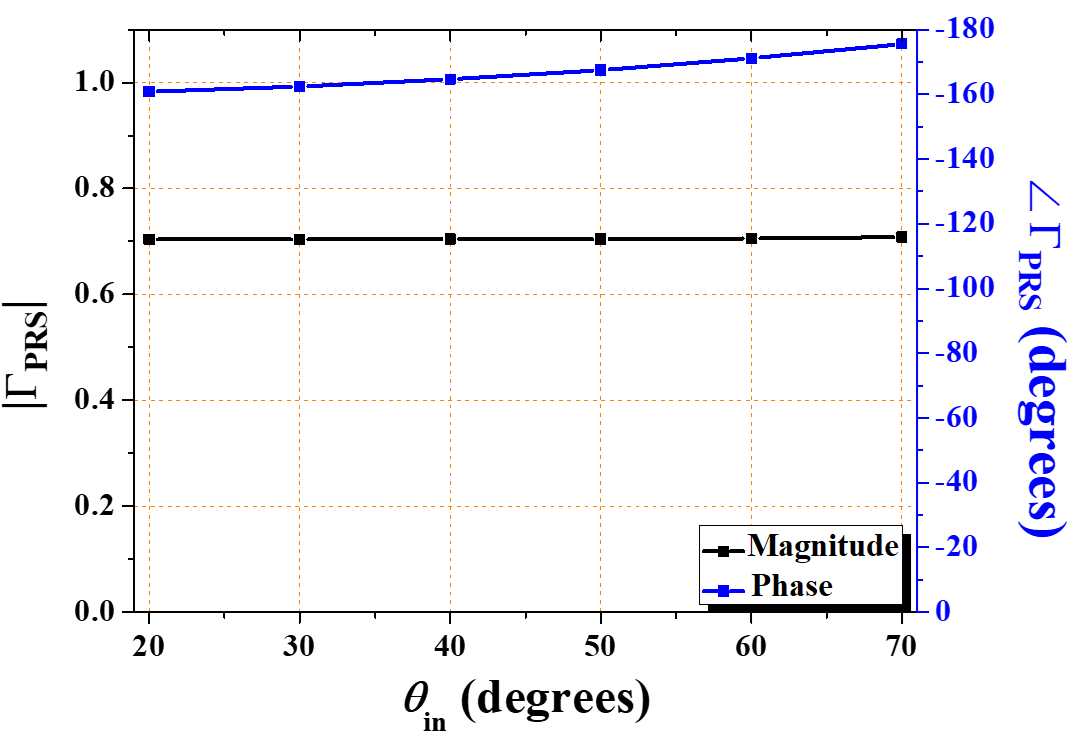}}
    \caption{Plot of the Floquet mode simulation results depicting the transmittance characteristics of the angle-independent PRS.}
    \label{fig:S11_PRS.PNG}
\end{figure}

To first address the angular dependence of the PRS, $\Gamma_{\text{PRS}}$ is evaluated as a function of the propagation angle, $\theta_{\text{in}}$, using the transverse equivalent circuit model (see the inset of Fig.~\ref{fig:ConceptLWAFig.PNG}(a)), expressed as~\cite{shaham2024Arxiv},
\begin{equation}
\Gamma_\text{PRS}=\frac{\rho_0+\rho_1k_{z,o}k_o^{-1}+\rho_2(k_{z,o}k_o^{-1})^2} {\delta_0+\delta_1k_{z,o}k_o^{-1}+\delta_2(k_{z,o}k_o^{-1})^2+\delta_3(k_{z,o}k_o^{-1})^3}
\label{eq:Gamma_PRS}
\end{equation}
where $k_{z,o}=k_o\text{cos}(\theta_{\text{in}})$ and the remaining coefficients are defined as~\cite{shaham2024Arxiv},
\begin{subequations}
    \begin{align}
        {\rho_0} &= \frac{j}{q}[\xi_\text{bot}(\xi_\text{top}\xi_\text{mid}-2\xi)-\xi(\xi_\text{top}-\xi_\text{bot})], \\
        {\rho_1} &= \xi_\text{mid}(\xi_\text{top}-\xi_\text{bot}), \\
        {\rho_2} &= jq\xi_\text{mid}-ju(\xi_\text{top}+\xi_\text{bot}), \\
        \delta_0 &=-\rho_0, \\
        \delta_1 &=\xi_\text{mid}(\xi_\text{top}+\xi_\text{bot})-2\xi, \\
        \delta_2 &= jq\xi_\text{mid}+ju(\xi_\text{top}+\xi_\text{bot}), \\
        {\delta_3} &=-2qu
    \end{align}
    \label{eq:Currents}
\end{subequations}
\noindent Here, $\xi_\text{top}$, $\xi_\text{top}$, and $\xi_\text{top}$ represent the normalized admittance values in each admittance surfaces forming the PRS. Particularly, they are defined as
\begin{subequations}
    \begin{align}
        \xi_\text{top} &= 1+jqY_\text{top}\eta_o, \\
        \xi_\text{mid} &=2+jqY_\text{mid}\eta_o, \\
        \xi_\text{bot} &=1+jqY_\text{bot}\eta_o,
    \end{align}
    \label{eq:Currents}
\end{subequations}
\noindent where $\xi=1+pq$. Based on Eq.~\eqref{eq:Gamma_PRS}, the values of $\xi_\text{top}$, $\xi_\text{mid}$, and $\xi_\text{bot}$ are optimized to achieve an angle-independent reflection coefficient. Specifically, a particle swarm optimization is used to optimize $\xi_\text{top}$, $\xi_\text{mid}$, and $\xi_\text{bot}$ such that the magnitude of $\Gamma_\text{PRS}$ is fixed at 0.7 for $\theta_{\text{in}} \in \{20^\circ, 70^\circ\}$. The optimized admittance surface values are $Y_\text{top,PRS}=-0.0033 \Omega^{-1}$, $Y_\text{mid,PRS}=-0.0059 \Omega^{-1}$, and $Y_\text{bot,PRS}=-0.0179 \Omega^{-1}$. Fig.~\ref{fig:S11_PRS.PNG} shows the resulting reflection coefficient as a function of $\theta_{\text{in}}$, obtained using Floquet mode simulations in Ansys HFSS. As shown, the magnitude of $\Gamma_\text{PRS}$ remains close to the target value of 0.7, while the phase remains relatively constant over the specified range of angles.

In addition, to mitigate the open-stopband effect, we invoke the theory of Fourier optics which states that an aperture field and the corresponding Fraunhofer radiation form a Fourier transform pair. In light of this theory, a passive BMS is placed on top of the PRS (see Fig.\ref{fig:ConceptLWAFig.PNG}) to convert the fundamental-mode radiation from the PRS into the $n^\text{th}$ spatial harmonic. For this purpose, two types of unit cells are arranged to provide fixed transmission phases of $+160^\circ$ and $-20^\circ$, with near-unity transmission magnitude, thereby creating a binary transmission profile expressed as
\begin{equation}
S_\text{21,bin} = \text{sgn}\left(\cos\left(\frac{2\pi y}{P}\right)\right)e^{-j20^\circ}, \label{eq:BinaryPattern}
\end{equation}
\noindent where $\text{sgn}$ denotes the signum function, and $P$ represents the periodicity of the binary pattern, defined as~\cite{abdo2018IEEE},
\begin{equation}
P = \frac{2\pi}{\left|k_o\sin\theta_a - k_o\sin\theta_b\right|}.
\label{eq:Periodicity}
\end{equation}
Here, $\theta_a$ and $\theta_b$ are set to $50^\circ$ and $0^\circ$, respectively. These angles are chosen to ensure that only one Fourier component, which corresponds to one propagating mode, resides within the visible region when the propagation angle of the guided mode varies from $20^\circ$ to $70^\circ$. In this way, when the radiation angle (i.e., $\theta_{\text{in}}$) from the PRS changes due to the TRM around $50^\circ$, the output angle from the BMS can also change around broadside, thereby overcoming the open-stopband effect. 

\begin{figure}[t]
    \centerline{\includegraphics[width=\columnwidth]{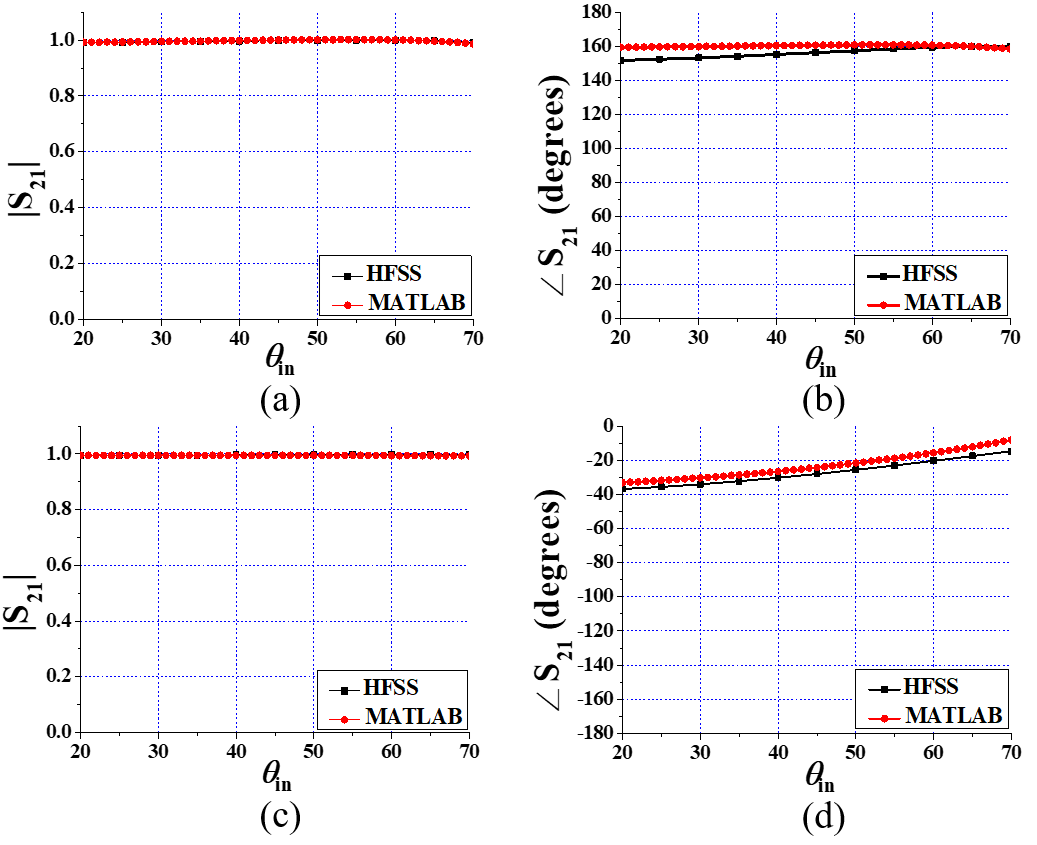}}
    \caption{Comparison of the transmittance of the BMS as obtained from HFSS simulations and MATLAB calculations. (a) Magnitude of the $160^\circ$ BMS, (b) Phase of the $160^\circ$ BMS, (c) Magnitude of the $-20^\circ$ BMS, and (d) Phase of the $-20^\circ$ BMS.}
    \label{fig:BMS_S21.PNG}
\end{figure}

Finally, to ensure that the binary transmission profile in Eq.~\eqref{eq:BinaryPattern} also remains consistent across all the incident angles within $\{20^\circ,70^\circ\}$, the angular dependence of the transmission coefficient, $T_\text{BMS}$, is evaluated as~\cite{shaham2024Arxiv},
\begin{equation} T_\text{BMS} = \frac{\tau_1k_{z,o}k_o^{-1} + \tau_3(k_{z,o}k_o^{-1})^3} {\delta_0 + \delta_1k_{z,o}k_o^{-1} + \delta_2(k_{z,o}k_o^{-1})^2 + \delta_3(k_{z,o}k_o^{-1})^3},
\label{eq:T_BMS}
\end{equation}
\noindent where $\tau_1 = 2\xi$ and $\tau_3 = 2q(2u - q)$.
Based on Eq.~\eqref{eq:T_BMS}, the admittance surfaces are optimized for full transmission, with the transmission phase set to either $160^\circ$ and $-20^\circ$ for $\theta_{\text{in}}\in\{20^\circ,70^\circ\}$. For a transmission phase of $160^\circ$, the corresponding admittance surfaces are calculated as $Y_\text{top,+160}=-0.0079 \Omega^{-1}$, $Y_\text{mid,+160}=-0.0323 \Omega^{-1}$, and $Y_\text{bot,+160}=-0.0079 \Omega^{-1}$. On the other hand, for a transmission phase of $-20^\circ$, the calculated admittance surfaces are $Y_\text{top,-20}=-0.001\Omega^{-1}$, $Y_\text{mid,-20}=0.0048\Omega^{-1}$, and $Y_\text{bot,-20}=-0.001\Omega^{-1}$. It is noted that, similar to before, these two unit cells are modeled as a cascade of three admittance surfaces, separated by a 1.575-mm thick Rogers 5880 substrate. Fig.~\ref{fig:BMS_S21.PNG} plots the optimized magnitude and phase variation with respect to the incident angles. As seen, the magnitude of $T_\text{BMS}$ is observed to approach near unity, and the phase also remains relatively constant. These findings suggest promising alignment with theoretical expectations, as will become evident through full-wave demonstrations presented in the following section. 

\section{Full-wave demonstration}
\begin{figure}[]
    \centerline{\includegraphics[width=0.95\columnwidth]{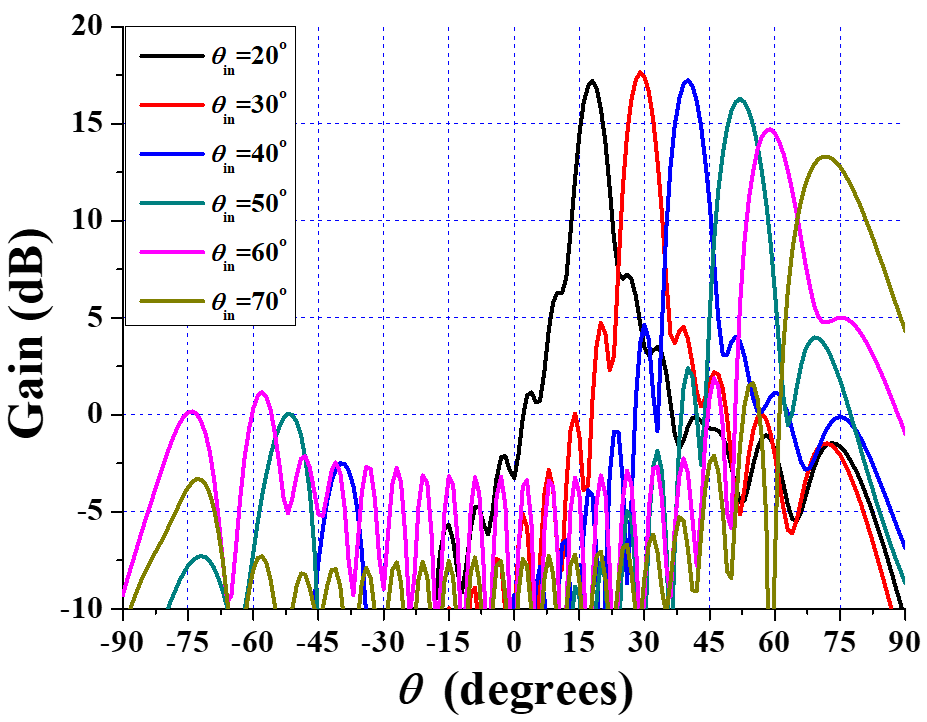}}
    \caption{Full-wave radiation patterns (gains) at various radiation angles for the fundamental mode radiation.}
    \label{fig:FundaModeRad7.PNG}
\end{figure}

%\subsection{Fundamental-mode beam steering}
To validate the proposed concept, an initial investigation is conducted by simulating the LWA depicted in Fig.~\ref{fig:ConceptLWAFig.PNG} in the absence of the BMS via ANSYS HFSS. In this simulation, the waveguide length is set to 10$\lambda_o$, and the TRM and PRS are modeled as the impedance boundaries to minimize computational costs. Without the BMS, it is expected that the PRS radiates the fundamental mode that propagates at $\theta_{\text{in}}$. By applying a bias voltage across the TRM, the reflective phase can be adjusted and alter $\theta_{\text{in}}$. Consequently, although our radiative aperture (i.e. the PRS) is entirely passive, the radiated beam can be dynamically steered. This is illustrated in Fig.~\ref{fig:FundaModeRad7.PNG} from which it is seen that the fundamental mode radiates at various $\theta_{\text{in}}$, which aligns well with the propagation angle within the LWA. It should be noted that Ohadi \textit{et al.} have recently proposed a tunable LWA that also allows dynamic steering of the fundamental mode by varying the propagation angle of a guided mode~\cite{ohadi2021IEEE}. In their work, however, two tunable metasurfaces were required, while the proposed work only utilizes one surface, thereby further simplifying the design architecture and biasing.

Over the beam-steering range of $20^\circ$ to $70^\circ$, the gain experiences a degradation of approximately 4 dB. This drop would be substantially larger if the PRS were not optimized for angle-independence, as non-optimized surfaces typically fail to maintain a consistent transmissive magnitude at oblique incidences. 

%It is noted the gain experiences a degradation of $\sim$4 dB across the beam-steering range of 20$^\circ$ to 70$^\circ$. If the PRS is not optimized to exhibit the angle-independent characteristics, the gain degradation would be substantially more pronounced due to their inability to sustain consistent transmissive magnitude at other oblique incident angles. While Fig.~\ref{fig:FundaModeRad7.PNG} highlights the advantage of the proposed concept in achieving beam-steering with simple biasing, it fails to overcome the open-stopband effect, as the fundamental mode is being radiated. In the next subsection, we show the case of beam steering across broadside by incorporating the BMS.

%In the next subsection, we incorporate the BMS to enable beam steering across broadside.
%validating the underlying concept of the proposed LWA.

\begin{figure}
    \centerline{\includegraphics[width=0.95\columnwidth]{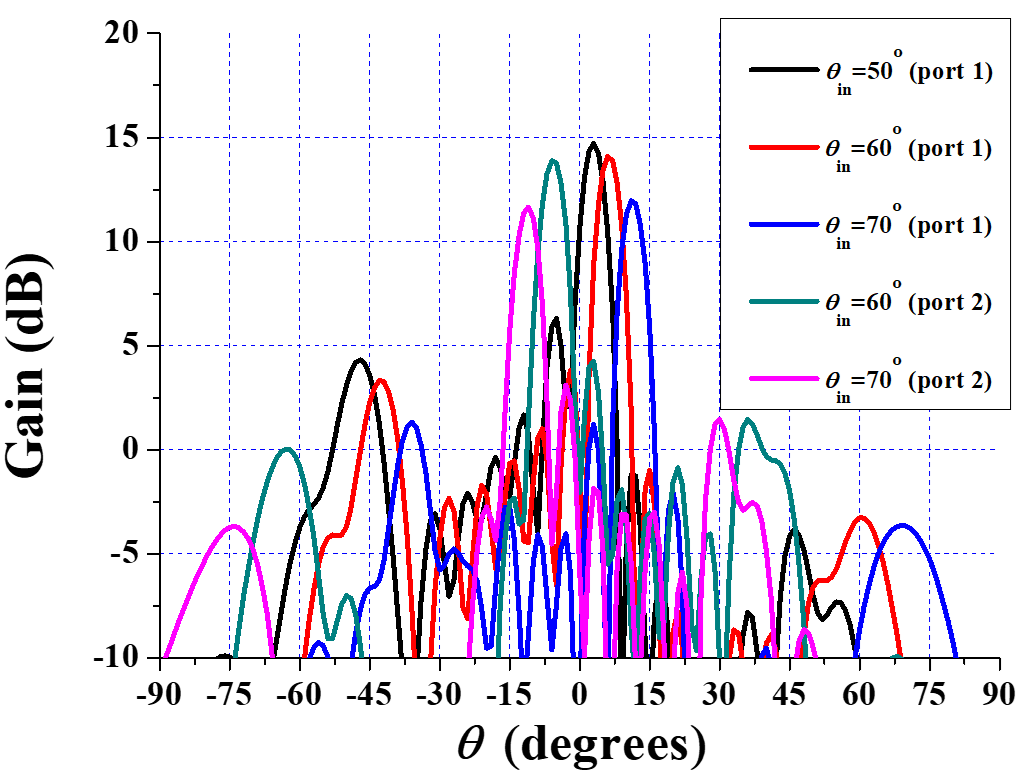}}
    \caption{Full-wave radiation patterns (gains) at various radiation angles. In this context, "port 1" denotes the scenario in which only port 1 is excited, with port 2 assumed to be perfectly matched. Conversely, "port 2" represents the complementary scenario, where only port 2 is excited under the same assumption.}
    \label{fig:LWARad4.PNG}
\end{figure}

%\subsection{Beam steering across broadside}
Although Fig.~\ref{fig:FundaModeRad7.PNG} demonstrates that our simple biasing scheme enables beam steering, it does not address the open-stopband effect because the fundamental mode is still being radiated. To resolve the open-stopband problem, we now consider the case where the BMS is incorporated, with its transmission profile defined in Eq.~\eqref{eq:BinaryPattern}. Specifically, the BMS is placed 1 mm above the PRS, and a secondary excitation port is introduced at the opposite end of the waveguide to extend the range of propagation angle from $-70^\circ$ to $+70^\circ$. It is reminded that the period of the binary pattern is chosen such that the LWA would radiate at broadside when the propagation angle corresponds to $50^\circ$. Fig.~\ref{fig:LWARad4.PNG} shows the full-wave simulation results of the proposed LWA with the BMS. As shown, when the propagation angle corresponds to 50$^\circ$, the maximum radiation occurs at $\theta_{\text{out}}=1^\circ$. Furthermore, as the propagation angle changes around 50$^\circ$, the radiation angle correspondingly scans across broadside, effectively eliminating the open-stopband effect.

%Owing to the angle-independent characteristics of the PRS and the BMS, the observed gain resulting from an incident angle of 70$^\circ$ exhibits minimal gain degradation. Specifically, it is noted that LWA utilizes a simplified biasing scheme to steer the beam by varying the reflective phase of the TRM, rather than directly modulating the radiative aperture.

\section{Conclusion}
This work presents a route to achieving dynamic beam steering at a fixed frequency using a novel metasurface-aided leaky-waveguide antenna that allows simplified biasing scheme. The design incorporates a tunable reflective metasurface (TRM) within the waveguide, which dynamically controls the propagation angle of a guided mode by varying its reflective phase. Simplified biasing is achieved by connecting all the unit cells of the TRM in parallel. For radiation, the partially reflective surface (PRS) is designed with partial transmission and angle-independent characteristics, ensuring efficient and consistent fundamental-mode radiation. Additionally, the binary metasurface (BMS) positioned above the PRS addresses the open-stopband problem. The feasibility of the proposed concept has been demonstrated through full-wave simulations using Ansys HFSS. Future efforts will focus on the physical realization of the proposed LWA, including experimental validation and further optimization for practical applications in dynamic beamforming technologies.

\section*{Acknowledgments}
This work was supported by the National Research Foundation of Korea (NRF) grants funded by the Korea government (MSIT) (RS-2024-00341191 and RS-2024-00343372).

\bibliographystyle{ieeetr}
\bibliography{SambongJang_Refs}

\end{document}